\documentclass[aps,prb,superscriptaddress,amsmath,amssymb,preprint]{revtex4-1}

\usepackage[utf8x]{inputenc}
\usepackage[english]{babel}
\usepackage{amsmath}
\usepackage{amssymb}
\usepackage{upgreek}
\usepackage{graphicx}
\usepackage{lmodern}
\usepackage{float}
\usepackage{siunitx}

\renewcommand{\bm}[1]{\boldsymbol{\mathbf{#1}}}
\usepackage{color}

\usepackage[colorlinks=true]{hyperref}

\definecolor{red}{rgb}{0.8,0,0}
\definecolor{JGUred}{rgb}{0.8,0,0}
\definecolor{JGUlightred}{rgb}{0.9,0.4,0.5}
\definecolor{JGUdarkgrey}{rgb}{0.3,0.3,0.3}
\definecolor{JGUgrey}{rgb}{0.7,0.7,0.7}
\definecolor{JGUlightgrey}{rgb}{0.85,0.85,0.85}
\definecolor{JGUblue}{rgb}{0.0157,0.35294,0.58038}
\definecolor{green}{rgb}{0.4,0.6,0.1}

\graphicspath{{./Figures/Fig1/}{./Figures/Fig2/}{./Figures/Fig3/}{./Figures/Fig4/}{./Figures/Fig5/}{./Figures/Fig6/}{./Figures/Fig7/}}

\begin{document}

\title{Impact of electromagnetic fields and heat on spin transport signals in Y\textsubscript{3}Fe\textsubscript{5}O\textsubscript{12}}

\author{Joel Cramer}
\affiliation{Institute of Physics, Johannes Gutenberg-University Mainz, 55099 Mainz, Germany}
\affiliation{Graduate School Materials Science in Mainz, 55128 Mainz, Germany}

\author{Lorenzo Baldrati}
\affiliation{Institute of Physics, Johannes Gutenberg-University Mainz, 55099 Mainz, Germany}

\author{Andrew Ross}
\affiliation{Institute of Physics, Johannes Gutenberg-University Mainz, 55099 Mainz, Germany}
\affiliation{Graduate School Materials Science in Mainz, 55128 Mainz, Germany}

\author{Mehran Vafaee}
\affiliation{Institute of Physics, Johannes Gutenberg-University Mainz, 55099 Mainz, Germany}

\author{Romain Lebrun}
\affiliation{Institute of Physics, Johannes Gutenberg-University Mainz, 55099 Mainz, Germany}

\author{Mathias Kl\"aui}
\email{Klaeui@uni-mainz.de}
\affiliation{Institute of Physics, Johannes Gutenberg-University Mainz, 55099 Mainz, Germany}
\affiliation{Graduate School Materials Science in Mainz, 55128 Mainz, Germany}

\date{\today}

	\begin{abstract}		
		Exploring new strategies to perform magnon logic is a key requirement for the further development of magnon-based spintronics.
		In this work, we realize a three-terminal magnon transport device to study the possibility of manipulating magnonic spin information transfer in a magnetic insulator via localized magnetic fields and heat generation.
		The device comprises two parallel Pt wires as well as a Cu center wire that are deposited on the ferrimagnetic insulator Y\textsubscript{3}Fe\textsubscript{5}O\textsubscript{12}.
		While the Pt wires act as spin current injector and detector, the Cu wire is used to create local magnetostatic fields and additional heat, which impact both the magnetic configuration and the magnons within the Y\textsubscript{3}Fe\textsubscript{5}O\textsubscript{12} below.
		We show that these factors can create a non-local signal that shows similar features as compared to an electrically induced magnon flow.
		Furthermore, a modulation of the spin transport signal between the Pt wires is observed, which can be partly explained by thermally excited spin currents of different polarization.
		Our results indicate a potential way towards the manipulation of non-local magnon signals, which could be useful for magnon logic.
	\end{abstract}

\maketitle

\section{Introduction}

The ultimate goal of magnonics\cite{Kruglyak2010,Serga2010,Lenk2011}, a research field that investigates spin wave phenomena, is the effective control and manipulation of spin wave propagation and detection in magnetic materials.
This involves the implementation of magnon circuitry that is able to perform logic operations, i.e. to realize magnon based computing\cite{Khitun2010,Chumak2015}.
Potential advantages of this approach over conventional, charge-based concepts are the possibility to encode information in both the amplitude and phase of a spin wave for coherent logic\cite{Chumak2015,Tserkovnyak2017} and furthermore an improved energy efficiency\cite{Khitun2010}, in particular in ferroic insulators.
The latter results from the absence of Joule heating during spin wave motion and low damping parameters when using magnetic insulators as a spin wave conduit\cite{Serga2010}.

To date, theoretical\cite{Kostylev2005,Klingler2014,Braecher2018} and experimental\cite{Chumak2014,Wagner2016,Fischer2017,Balinskiy2018} studies regarding magnon logic operations have mainly focused on schemes based on coherent spin waves effects, effectively exploiting interference phenomena\cite{Chumak2015,Tserkovnyak2017}.
This includes, for instance, devices like the all-magnon transistor\cite{Chumak2014}, a logic majority gate\cite{Fischer2017} or an analog magnon adder\cite{Bracher2018}.
Lately, logic operations based on thermal, incoherent magnons as information carriers have gained increased interest in the course of non-local magnon transport experiments\cite{Cornelissen2015,Goennenwein2015,lebrun2018tunable}.
Through the direct and inverse spin Hall effect (SHE)\cite{Sinova2015}, which appear in conductors with strong spin-orbit interaction (SOI), information transport in magnetic insulators via thermal magnons can be excited and detected electrically, interfacing magnonics and electronics.
Among others, it has been successfully demonstrated that the linear superposition of diffusive magnon currents in insulators can be used to implement a majority gate\cite{Ganzhorn2016}.
Furthermore, multi-terminal devices that exhibit transistor-like behavior have been realized\cite{Cornelissen2018,Wimmer2018}.
The latter rely on the local manipulation of the magnon chemical potential\cite{Cornelissen2016b} and thus the magnon conductivity, which is achieved by the SHE induced injection of additional magnons through a heavy metal gate.
However, applying a charge current to such a gate concurrently results in local heating and generates a magnetostatic field (Oersted field), which also might affect magnon propagation within the insulator.
To study these effects exclusively, without any interference due to externally injected magnon currents, one would need to replace the heavy metal gate by a normal metal with weak SOI and, hence, a negligible SHE.

In this work, a three-terminal non-local magnon transport device including a Cu-based gating structure is implemented to investigate the modulation of magnon propagation signals in magnetic insulators via localized magnetostatic fields and heat.
Magnon transport measurements in rotating magnetic fields of varying amplitude show that these perturbing forces affect the non-local signal and furthermore generate an additional voltage response with similar features as that of magnons induced by the spin Hall effect.

\section{Experimental details}

\begin{figure}[!b]
	\centering
	\includegraphics[width=8.6 cm]{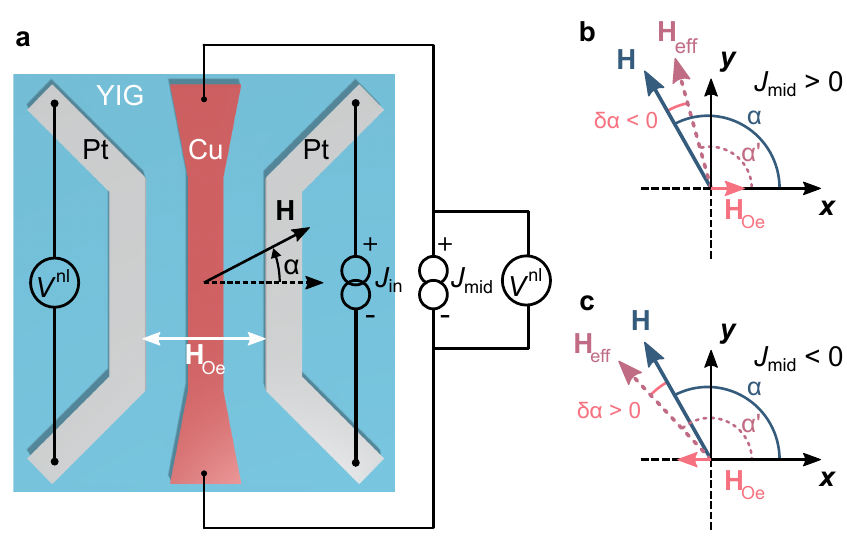}
	\caption{(a) Schematic view of the fabricated non-local device structure.
		Besides two parallel Pt wires (injector and detector), a third Cu wire is positioned in the center of the device.
		Depending on the performed experiment, the electrical wiring changes.
		(b),(c) Illustration of the effective field $\bm{H}_\mathrm{eff} = \bm{H} + \bm{H}_\mathrm{Oe}$ and thus the angular shift $\delta \alpha$ for different charge current polarities applied to the Cu wire.
		Adapted from [\onlinecite{cramer2018phd}].} 
	\label{fig:ptcupt0}
\end{figure}

A schematic of the implemented multi-terminal device is shown in Fig.~\ref{fig:ptcupt0}a, depicting a non-local magnon transport structure that comprises three metallic nanowires on a magnetic insulator.
Regarding the latter, a commercially available, single crystalline Y\textsubscript{3}Fe\textsubscript{5}O\textsubscript{12} (YIG)\cite{Serga2010} film with a thickness of $d_\mathrm{YIG} = \SI{150}{\nano\meter}$ and with (111) surface orientation is used.
The outer wires (\SI{250}{\nano\meter} width and \SI{1.1}{\micro\meter} center-to-center distance) are made of Pt ($d_\mathrm{Pt} = \SI{7.5}{\nano\meter}$), while the center strip (\SI{250}{\nano\meter} width) consists of a \SI{15}{\nano\meter} copper layer capped by \SI{5}{\nano\meter} of Al to protect it from oxidation.
The nanowires are patterned in a multi-step lift-off process including electron beam lithography and metal deposition via magnetron sputtering.

In the device, the Pt wires are used to both inject and detect magnonic spin currents in the YIG through the direct and inverse SHE\cite{Cornelissen2015,Goennenwein2015}.
Considering first the excitation of spin currents, the application of a charge current $J_\mathrm{in}$ to one of the wires (injector) results in a spin-dependent, transverse deflection of electrons due to the SHE so that, eventually, a spin accumulation $\bm{\mu}_\mathrm{s}$ builds up at the Pt/YIG interface.
The polarization vector of this spin accumulation is perpendicular to the wire and, depending on the magnetic orientation of the YIG, magnons are either created or annihilated\cite{Bender2012}.
As a result, an imbalance of the magnon population in the YIG is induced and a diffusive magnon spin current is flowing\cite{Zhang2012}.
Concurrently, a thermally excited magnon flow generated by the spin Seebeck effect (SSE) is present\cite{Uchida2016,Ganzhorn2017} due to the Joule heating from $J_\mathrm{in}$.
With regard to the detection of magnonic spin currents, these are partially absorbed by the second Pt stripe (detector) and reconverted to a detectable charge signal via the inverse SHE\cite{Cornelissen2015}.

As mentioned above, the novelty of our device is the Cu center strip that is used here to study the modulation of the magnon transport signal between injector and detector via the generation of local heat and magnetostatic Oersted fields $\bm{H}_\mathrm{Oe}$ supplied by the charge current $J_\mathrm{Cu}$ (see Fig.~\ref{fig:ptcupt0}a).
Note that, in contrast to Pt, Cu does not exhibit a sizable SHE/ISHE due to its weak SOI\cite{Wang2014} such that $J_\mathrm{Cu}$ does not result in a further spin current pumped into the YIG\cite{Cornelissen2018,Wimmer2018}, which would interfere with the effects studied here.

For the full characterization of the device and identification of various signal contributions, different circuitry configurations were implemented, see Fig.~\ref{fig:ptcupt0}a.
Spin transport between the outer Pt stripes (Pt~$\rightarrow$~Pt) was studied by connecting the right Pt wire (injector) to a current source supplying the DC charge current $J_\mathrm{Pt} = \pm\SI{250}{\micro\ampere}$ ($j_\mathrm{Pt} \approx \SI{1.3E11}{\ampere\per\meter\squared}$) required for the electrical (SHE) or thermal (SSE) excitation of magnons in the YIG.
The inverse SHE voltage drop at the detector (left Pt stripe) was picked up by a nanovoltmeter.
The Cu center wire (modulator) was connected to a second current source, supplying a DC charge current $J_\mathrm{Cu}$ of up to $\pm\SI{1}{\milli\ampere}$ ($j_\mathrm{Cu} \leq \pm \SI{2.29E11}{\ampere\per\meter\squared}$).
In a second configuration, the direct response of the detector towards charge currents applied to the modulator (Cu~$\rightarrow$~Pt) was checked, for which the Pt injector was left unbiased ($J_\mathrm{Pt} = \SI{0}{\ampere}$).
Eventually, to verify that the Cu modulator reveals no spin-charge interconversion, it was connected to the nanovoltmeter while applying a charge current to the injector (Pt~$\rightarrow$~Cu).
In the following, we show field and angular-dependent measurements performed by sweeping an external field $\bm{H}$ or by rotating the sample in a static field (angle $\alpha$, see Fig.~\ref{fig:ptcupt0}a).
All measurements were conducted at room temperature.

Note that when implementing a DC measurement scheme as described above, electrically (SHE) and thermally (SSE) induced spin signals typically can be extracted by considering either the difference or the sum of the non-local voltages obtained for positive and negative charge currents applied to the injector\cite{Thiery2018}.
In this study, such a simple distinction is not always applicable so that we use a generalized notation
\begin{align}
V^\mathrm{nl}_\mathrm{\Delta} &= \left[ V_\mathrm{nl} \left( + J \right) - V_\mathrm{nl} \left( - J \right) \right] / 2., \\
V^\mathrm{nl}_\mathrm{\Sigma} &= \left[ V_\mathrm{nl} \left( + J \right) + V_\mathrm{nl} \left( - J \right) \right] / 2. \label{eq:vsigma}
\end{align}
Here, $J$ is the charge current applied to either the Pt injector ($J_\mathrm{Pt}$) or the Cu wire ($J_\mathrm{Cu}$).
If an AC measurement scheme was used as in Ref.~\onlinecite{Cornelissen2015}, $V^\mathrm{nl}_\mathrm{\Delta}$ ($V^\mathrm{nl}_\mathrm{\Sigma}$) would correspond to the $1\omega$ ($2\omega$) signal.

\section{Results and discussion}

\subsection{Device functionality}

\begin{figure}[!t]
	\centering
	\includegraphics[width=8.6 cm]{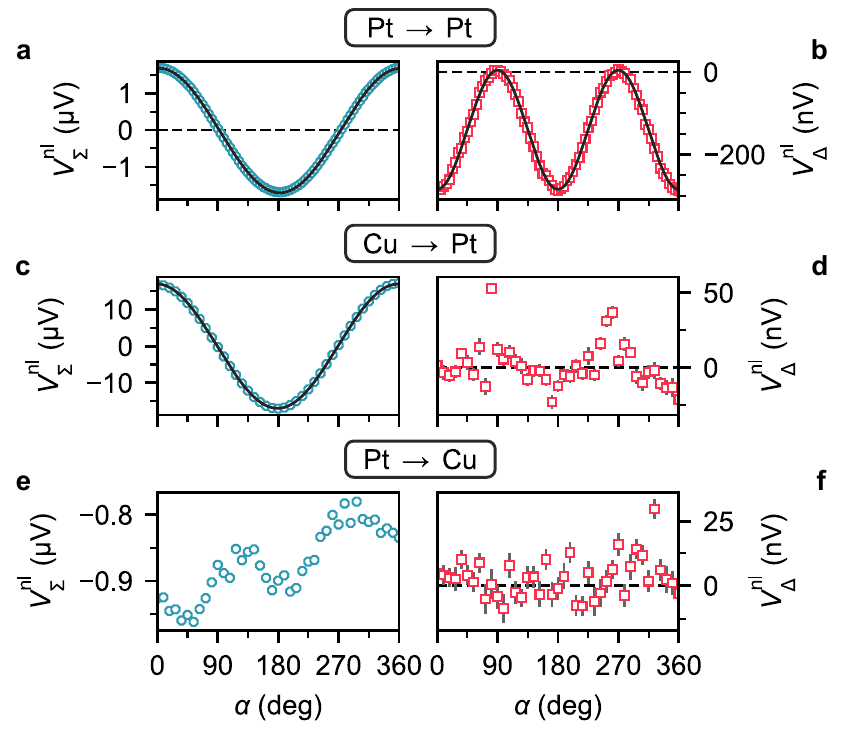}
	\caption{Non-local voltages due to (a),(c),(e) thermal ($V^\mathrm{nl}_\mathrm{\Sigma}$) and (b),(d),(f) electrical ($V^\mathrm{nl}_\mathrm{\Delta}$) excitation of spin currents as a function of  the in-plane field direction $\alpha$.
		In (a),(b) data obtained for Pt~$\rightarrow$~Pt is shown, whereas (c),(d) display the results for Cu~$\rightarrow$~Pt and (e),(f) for Pt~$\rightarrow$~Cu.
		In all graphs, solid lines give the results of corresponding fit functions.
		The external field amplitude is $H = \SI{750}{Oe}$, sufficient to fully saturate the YIG.
		If visible, error bars give the standard error.
		Adapted from [\onlinecite{cramer2018phd}].	}
	\label{fig:ptcupt1}
\end{figure}

First, we checked the functionality of the non-local device and especially the absence of any spin-charge conversion in the Cu wire.
The Pt~$\rightarrow$~Pt configuration with no charge current applied to the Cu wire shown in Fig.~\ref{fig:ptcupt1}a,b
reveals the angular dependences expected for thermally [$V^\mathrm{nl}_\mathrm{\Sigma} \propto \cos \left(\alpha\right)$] and electrically [$V^\mathrm{nl}_\mathrm{\Delta} \propto \cos ^2 \left(\alpha\right)$] excited magnon currents, in agreement with the spin current excitation and detection model\cite{Zhang2012,Bender2012}.
These dependences arise from the (I)SHE symmetry, which yields maximum injection and detection signals when the YIG magnetization is parallel to the spin accumulation $\bm{\mu}_\mathrm{s}$, while being zero for perpendicular alignment.
The non-local voltage obtained for the Cu~$\rightarrow$~Pt configuration, i.e. the spin signal generated when injecting a charge current into the Cu modulator, is now shown in Fig.~\ref{fig:ptcupt1}c,d.
The current-induced Joule heat gives rise to the non-local SSE, as captured by $V^\mathrm{nl}_\mathrm{\Sigma}$, while $V^\mathrm{nl}_\mathrm{\Delta}$ fluctuates around zero.
This is expected as Cu does not exhibit a significant SHE\cite{Wang2014}.
The higher amplitude of $V^\mathrm{nl}_\mathrm{\Sigma}$ in Fig.~\ref{fig:ptcupt1}c as compared to that in Fig.~\ref{fig:ptcupt1}a is due to the smaller wire distance and stronger Joule heating ($\approx \SI{6.40}{\milli\watt}$ vs. $\approx \SI{0.73}{\milli\watt}$).
Moreover, the results depicted in Fig.~\ref{fig:ptcupt1}e,f (Pt~$\rightarrow$~Cu configuration) additionally corroborate the absence of spin-charge conversion in Cu.
$V^\mathrm{nl}_\mathrm{\Delta}$ fluctuates around zero, while $V^\mathrm{nl}_\mathrm{\Sigma}$ shows a finite, oscillating and slightly increasing voltage.
The geometry of this last signal, however, does not correspond to the one of the SSE and therefore must be of different origin (e.g. conventional Seebeck effect and fluctuating sample temperature in our thermally non-isolated setup).

These results demonstrate the current-induced spin transport between the outer Pt stripes and the absence spin-charge conversion within the Cu wire.
The latter information is crucial for the following discussion.

\subsection{Spin signal excitation via the Cu wire}

In this section, we now consider the effect of magnetostatic Oersted fields at the modulator, which are generated by $J_\mathrm{Cu}$\cite{coey2010magnetism}:
\begin{equation}
\oint \bm{H}_\mathrm{Oe} \mathrm{d}\bm{s} = J_\mathrm{Cu},\label{eq:ampere}
\end{equation}
where $\oint \mathrm{d}\bm{s}$ describes a closed integrating path around the Cu modulator ($xz$-plane).
Charge currents applied to the nanowires flow along the $y$-direction (see Fig.~\ref{fig:ptcupt0}) such that $\bm{H}_\mathrm{Oe}$ exclusively exhibits $x$- and $z$-components.
Here, only the $x$-component is of interest considering a relatively strong easy plane shape anisotropy of the thin YIG film.
According to Eq.~\ref{eq:ampere}, $J_\mathrm{Cu} >0$ results in a positive Oersted field underneath the Cu wire ($x$-component in $\alpha = \ang{0}$ direction), while $J_\mathrm{Cu}<0$ yields a negative component ($\alpha = \ang{180}$).
For the following discussions, it is useful to consider an effective field $\bm{H}_\mathrm{eff} = \bm{H} + \bm{H}_\mathrm{Oe}$ at an angle $\alpha^\prime = \alpha + \delta \alpha$, which locally acts on the YIG magnetization.
As sketched in Fig.~\ref{fig:ptcupt0}b,c, the angular shift $\delta \alpha $ is negative for $ \ang{0} < \alpha < \ang{180}$ and $J_\mathrm{Cu} > 0$, while being positive for $J_\mathrm{Cu} < 0$.
In the range of $ \ang{180} < \alpha < \ang{360}$, the  opposite occurs.

Finite element simulations (see Supporting Information) show that for the maximum current $J_\mathrm{Cu} = \pm \SI{1}{\milli\ampere}$ a field amplitude of $H_{\mathrm{Oe},x} \approx \pm \SI{14}{Oe}$ is induced at the wire center, directly at the YIG/Cu interface.
The external field applied during the angular-dependent measurements of Fig.~\ref{fig:ptcupt1}, however, was much larger ($H = \SI{750}{Oe}$) such that the additional torque exerted on the YIG magnetization by the Oersted field can be considered negligible.
Note that the magnitude of $\bm{H}_\mathrm{Oe}$ exhibits a strong spatial variation, so that the net effect of the Oersted field on the YIG magnetization configuration is hereafter discussed qualitatively rather than quantitatively, for which further simulations would be necessary.

\begin{figure}[!b]
	\centering
	\includegraphics[width=8.6 cm]{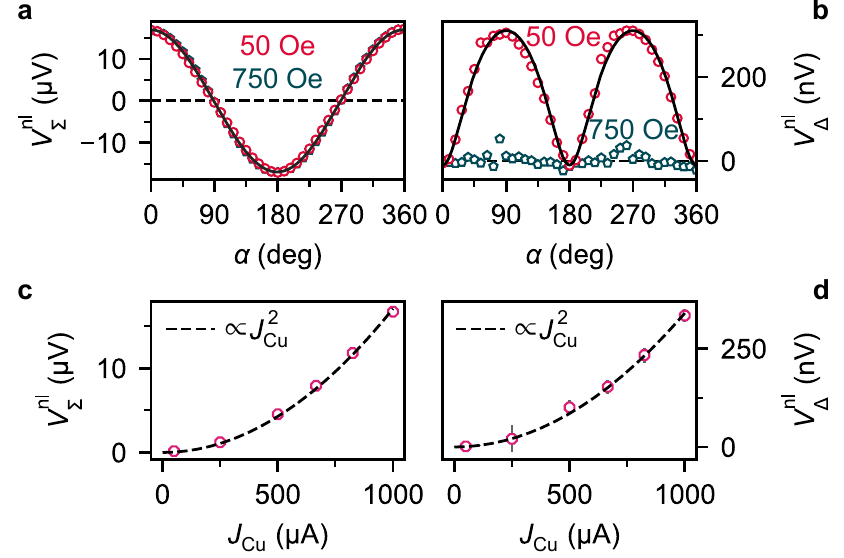}
	\caption{Angular dependence of (a) $V^\mathrm{nl}_\mathrm{\Sigma}$ and (b) $V^\mathrm{nl}_\mathrm{\Delta}$ recorded using the Cu~$\rightarrow$~Pt configuration.
		In each graph, the results obtained for low ($H = \SI{50}{Oe}$, red circles) and high ($H = \SI{750}{Oe}$, blue pentagons) external magnetic field amplitude are directly compared.
		Solid lines give the result of corresponding fit functions (please see main text).
		(c),(d) Current dependence ($J_\mathrm{Cu}$) of (c) the $V^\mathrm{nl}_\mathrm{\Sigma}$ amplitude and (d) the $V^\mathrm{nl}_\mathrm{\Delta}$ amplitude, recorded at $H = \SI{50}{Oe}$.
		The dashed line corresponds to a quadratic fit function.
		If visible, error bars are for (a),(b) the standard error and for (c),(d) errors of the fit function.
		Adapted from [\onlinecite{cramer2018phd}].	}
	\label{fig:ptcupt2}
\end{figure}

To probe the impact of the Oersted field, the angular-dependent measurements in the Cu~$\rightarrow$~Pt configuration were repeated at a reduced external field of $H = \SI{50}{Oe}$, which is enough for the YIG magnetization to follow the field direction (see Supporting Information).
As shown in Fig.~\ref{fig:ptcupt2}a, $\bm{H}_\mathrm{Oe}$ does not modulate $V^\mathrm{nl}_\mathrm{\Sigma}$, which stems from the conventional non-local SSE.
This appears reasonable, as $V^\mathrm{nl}_\mathrm{\Sigma}$ is even in the direction of $\bm{H}_\mathrm{Oe}$ and potential effects by the Oersted field are averaged out.
$V^\mathrm{nl}_\mathrm{\Delta}$, on the other hand, is odd in the direction of $\bm{H}_\mathrm{Oe}$ and a distinct, angular-dependent signal is recorded for $H = \SI{50}{Oe}$.
The symmetry of the signal does not agree with the one of electrically (SHE) injected spin currents, however it can be fitted by an adjusted function, yielding an amplitude of \SI[separate-uncertainty=true]{337 \pm 10}{\nano\volt}.
The definition of this function is discussed in the Supporting Information, based on the findings shown below.

\begin{figure}[!b]
	\centering
	\includegraphics[width=8.6 cm]{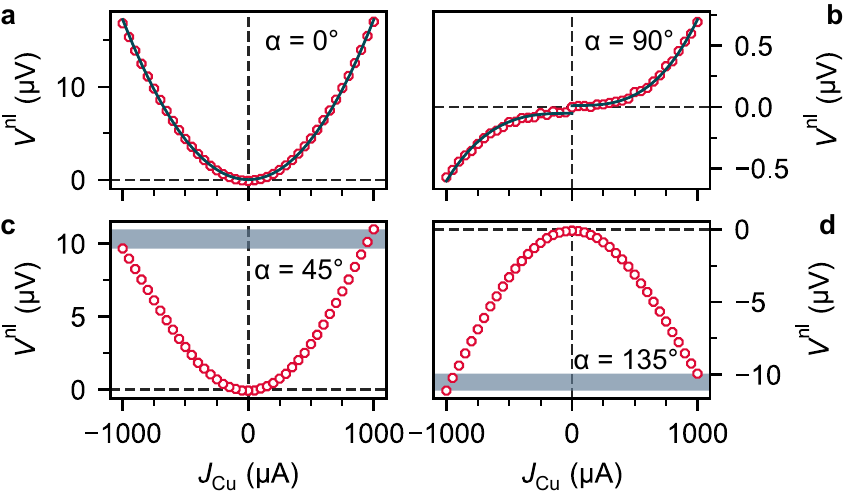}
	\caption{Raw voltage drop $V^\mathrm{nl}$ at the Pt detector as a function of $J_\mathrm{Cu}$ and at different magnetization directions:
		(a) $\alpha=\ang{0}$, (b) $\alpha=\ang{90}$, (c) $\alpha=\ang{45}$ and (d) $\alpha=\ang{135}$.
		The solid line in (a) corresponds to a quadratic fit function, whereas the one in (b) describes a cubic current dependence.
		The shaded areas in (c),(d) mark the voltage level asymmetries for positive and negative $J_\mathrm{Cu}$
		During the measurements, the external field was switched off after alignment of the magnetization (remanence).
		Error bars accounting for the standard error are smaller than the symbol size.
		Adapted from [\onlinecite{cramer2018phd}].	}
	\label{fig:ptcupt4}
\end{figure}

To demonstrate the origin of the $V^\mathrm{nl}_\mathrm{\Delta}$ signal, we measured the angular dependence for different $J_\mathrm{Cu}$ applied to the modulator.
Considering first $V^\mathrm{nl}_\mathrm{\Sigma}$ (Fig.~\ref{fig:ptcupt2}c), we see a quadratic current dependence as expected for spin currents generated by the SSE\cite{Thiery2018}.
Likewise, fitting the $V^\mathrm{nl}_\mathrm{\Delta}$ data yields a quadratic $J_\mathrm{Cu}$ dependence of the amplitude, see Fig.~\ref{fig:ptcupt2}d. 
While this indicates a thermal origin of $V^\mathrm{nl}_\mathrm{\Delta}$, one also has to bear in mind the direction of $\bm{H}_\mathrm{Oe}$ and the induced angular shift $\delta \alpha$ as a function of the applied charge current $J_\mathrm{Cu}$.
The largest $\delta\alpha$  is expected for $\alpha = \ang{90},\,\ang{270}$, where the $x$-component of the external field is zero.
For $\alpha = \ang{0},\ang{180}$, on the other hand, $\delta \alpha = \ang{0}$ is anticipated as $\bm{H}_\mathrm{Oe} \parallel \bm{M}_\mathrm{YIG}$.
To verify this behavior, the current dependence of the raw voltage $V^\mathrm{nl}$ is shown in Fig.~\ref{fig:ptcupt4} for different magnetization directions.
In these measurements, the external field was used to align $\bm{M}_\mathrm{YIG}$ at an desired angle $\alpha$ and switched off afterwards (remanent state of YIG) to observe the pure impact of the Oersted field.
Considering first $\alpha = \ang{0}$, one can see in Fig.~\ref{fig:ptcupt4}a that $V^\mathrm{nl}$ is a quadratic function of $J_\mathrm{Cu}$, as expected for thermally excited signals\cite{Thiery2018}.
The symmetry of the voltage furthermore verifies that the additional Oersted field has no effect.
Remarkably, for $\alpha = \ang{90}$ $V^\mathrm{nl}$ exhibits an asymmetric, cubic current dependence, see Fig.~\ref{fig:ptcupt4}b.
This asymmetry immediately demonstrates the importance of the Oersted field: the positive voltage for $J_\mathrm{Cu} > 0$ corresponds to a non-local SSE signal at $\alpha^\prime < \ang{90}$, while $J_\mathrm{Cu}<0$ yields $\alpha^\prime > \ang{90}$ and thus a negative SSE voltage.
The relevance of $\bm{H}_\mathrm{Oe}$ becomes further evident in Fig.~\ref{fig:ptcupt4}c,d, respectively, in which $\bm{M}_\mathrm{YIG}$ was aligned along $\alpha = \ang{45}$ and $\alpha = \ang{135}$.
The opposite voltage signs signify the reversed $x$-components of $\bm{M}_\mathrm{YIG}$ and thus the reversed polarization of magnonic spin currents flowing, whereas the inverted asymmetries regarding $J_\mathrm{Cu} \gtrless 0$ corroborate the Oersted field effect.
At $\alpha = \ang{45}$, a positive (negative) $J_\mathrm{Cu}$ applied to the Cu manipulator results in $\alpha^\prime < \ang{45}$ ($\alpha^\prime > \ang{45}$), yielding different amplitudes.
For $\alpha = \ang{135}$, similar considerations can be made.

As a first conclusion, we have shown that the finite voltage signal $V^\mathrm{nl}_\mathrm{\Delta}$, which appears at low or zero external fields applied, measures the difference between the $x$-components of thermally excited magnonic spin currents.
The ISHE in the Pt detector is sensitive to this variation, induced by the influence of the reversed Oersted fields.
Such fields can have significant impact at low external field amplitudes ($\delta \alpha \neq \ang{0}$), whereas $V^\mathrm{nl}_\mathrm{\Delta}$ becomes zero at large external fields due to $\bm{H}_\mathrm{eff} \simeq \bm{H}$ for the $J_\mathrm{Cu}$ amplitudes used in this work.

Besides the discussed asymmetries, this model also explains the cubic current dependence of $V^\mathrm{nl}$ on $J_\mathrm{Cu}$ in Fig.~\ref{fig:ptcupt4}b.
As mentioned before, the non-local SSE voltage is proportional to the Joule heat generated by the Cu wire ($\propto J_\mathrm{Cu}^2$) and to the $x$-component of the YIG magnetization $M^x_\mathrm{YIG} = M_\mathrm{YIG}\cdot \cos \left(\alpha^\prime\right)$:
\begin{equation}
V_\mathrm{nl} \propto \cos \left(\alpha^\prime\right) \cdot J_\mathrm{Cu}^2. \label{eq:thermaloerstedstuff}
\end{equation}
At $\alpha = \ang{90}$, this yields
\begin{equation}
V_\mathrm{nl} \propto \sin \left(\delta\alpha\right) \cdot J_\mathrm{Cu}^2 \approx \delta \alpha J_\mathrm{Cu}^2 \label{eq:thermaloerstedstuff2}
\end{equation}
for small $\delta\alpha$. 
Moreover, at this angle the $x$-component of the effective field is exclusively provided by $\bm{H}_\mathrm{Oe}$ such that, according to Eq.~\ref{eq:ampere}, $\delta\alpha \propto J_\mathrm{Cu}$ and
\begin{equation}
V_\mathrm{nl}^\mathrm{therm.} \propto J_\mathrm{Cu}^3. \label{eq:cubic}
\end{equation}
Among other things, the proportionality factor in Eq.~\ref{eq:cubic} is determined by the amplitude of the external field.

Taken all together, these results demonstrate that even when using a metal wire with negligible SOI, a finite $V^\mathrm{nl}_\mathrm{\Delta}$ can appear.
In contrast to a heavy metal injector, this voltage is not given by SHE induced spin currents, but results from different $x$-components of thermal spin currents generated in the presence of opposite Oersted fields.
Thus, when interpreting $V^\mathrm{nl}_\mathrm{\Delta}$ in general, multiple mechanisms need to be analyzed.

\subsection{Impact of heat and Oersted fields on spin transport signals}

\begin{figure}[!b]
	\centering
	\includegraphics[width=8.6 cm]{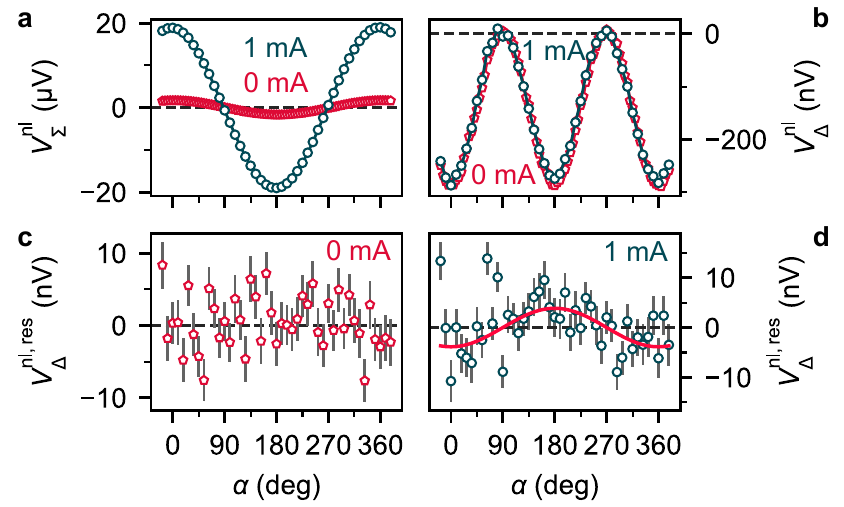}
	\caption{(a) $V^\mathrm{nl}_\mathrm{\Sigma}$ and (b) $V^\mathrm{nl}_\mathrm{\Delta}$ [Pt~$\rightarrow$~Pt configuration] as a function of $\alpha$, measured for two different charge currents applied to the Cu wire ($J_\mathrm{Cu} = \SI{0}{\milli\ampere},\,+\SI{1}{\milli\ampere}$).
	The amplitude of the external field is $H = \SI{750}{Oe}$.
	(c),(d) Calculated residuals of $V^\mathrm{nl}_\mathrm{\Delta}$ [difference of data in Fig.~\ref{fig:ptcupt5}b and fit function $V^\mathrm{nl}_\mathrm{\Delta} \propto \cos^2\left(\alpha\right)$] for (a) $J_\mathrm{Cu} = \SI{0}{\milli\ampere}$ and (b) $J_\mathrm{Cu} = +\SI{1}{\milli\ampere}$.
	Displayed error bars give the propagated standard error and fitting errors.
	In (d), the red solid line gives a phenomenological fit curve ($V^\mathrm{nl,res}_\mathrm{\Delta} \propto \cos\left(\alpha\right)$).
		Adapted from [\onlinecite{cramer2018phd}].}
	\label{fig:ptcupt5}
\end{figure}

As described in the introduction, the actual aim of this work is to modulate the spin information exchange between heavy metal wires by the heat and Oersted fields generated by the Cu center wire.
Focusing first on the impact of Joule heating, angular-dependent measurements implementing the Pt~$\rightarrow$~Pt configuration (see Fig.\ref{fig:ptcupt1}a,b) have been performed, once with and without a charge current applied to the Cu wire ($J_\mathrm{Cu} = \SI{0}{\milli\ampere},\SI{1}{\milli\ampere}$).
To suppress any influence of the Oersted field, the sample was rotated in an external field of $H = \SI{750}{Oe}$.
As shown in Fig.~\ref{fig:ptcupt5}a, the variation of $V^\mathrm{nl}_\mathrm{\Sigma}$ for different $J_\mathrm{Cu}$ is trivial:
Due to the additional heat, the signal amplitude is significantly enhanced.
This result emphasizes that, by definition, $V^\mathrm{nl}_\mathrm{\Sigma}$ accounts for thermally excited magnons that are generated near the Pt injector and the Cu modulator.
Instructive information on potential modulations of the spin transport mechanism between the Pt wires is thus only provided by $V^\mathrm{nl}_\mathrm{\Delta}$, for which the modulator contributions average out.
Therefore, $V^\mathrm{nl}_\mathrm{\Sigma}$ is disregarded in the following.

Figure ~\ref{fig:ptcupt5}b shows $V^\mathrm{nl}_\mathrm{\Delta}$ as a function of $\alpha$, for zero and finite $J_\mathrm{Cu}$.
At first sight, it seems that the charge current through the Cu wire yields no modification of the spin transport signal.
For both $J_\mathrm{Cu} = \SI{0}{\milli\ampere}$ and $J_\mathrm{Cu} = +\SI{1}{\milli\ampere}$ the expected angular dependence [$\cos^2\left(\alpha\right)$] is observed and similar signal amplitudes are obtained: $V^\mathrm{nl}_\mathrm{\Delta} \left(\SI{0}{\milli\ampere}\right) = \SI[separate-uncertainty=true]{-282 \pm 5}{\nano\volt}$ and $V^\mathrm{nl}_\mathrm{\Delta} \left(+\SI{1}{\milli\ampere}\right) = \SI[separate-uncertainty=true]{-280 \pm 4}{\nano\volt}$.
To nevertheless identify potential smaller deviations from the conventional spin transport signal, the experimental data was fitted by a correpsonding function and the resultant curve was subtracted from the data.
The calculated residuals are displayed in Fig.~\ref{fig:ptcupt5}c,d.
For zero applied charge current (Fig.~\ref{fig:ptcupt5}c), the residuals fluctuate around \SI{0}{\volt} without distinct angular dependence, as expected.
For $J_\mathrm{Cu} = +\SI{1}{\milli\ampere}$ (Fig.~\ref{fig:ptcupt5}d), however, a variation of  $V^\mathrm{nl,res}_\mathrm{\Delta}$ as a function of $\alpha$ appears.
Assuming the symmetry of the signal, the data is fitted phenomenologically by a cosine function, yielding a modulation amplitude of \SI[separate-uncertainty=true]{3.68 \pm 0.78}{\nano\volt}.
To check the unambiguity of this modulation, the \textit{reduced $\chi^2$} method\cite{hughes2010measurements} is applied, which allows one to evaluate whether the data is consistent with the proposed model or if the model is to be rejected.
For the cosine function, a reduced chi-squared of $\chi^2_\mathrm{red} \approx 0.98$ is calculated, implying that it describes the data well\cite{hughes2010measurements}.
The alternative model of no signal modulation (horizontal line) yields $\chi^2_\mathrm{red} \approx 2.71$ and thus can be rejected \cite{hughes2010measurements}.
One therefore can speculate whether the diffusive SSE current, whose polarization is determined by the rotating magnetization $\bm{M}_\mathrm{YIG}$, interacts with the electrically induced magnons propagating in the system\cite{Wimmer2018}.
To obtain a better understanding of this effect and to explain the observed angular dependence quantitatively, further experimental as well as theoretical work is required.
Finally, note that the Joule heat emitted by the Cu modulator entails a local reduction of the YIG magnetization due to the enhanced temperatures\cite{Gilleo1958}.
This can result in a generally decreased amplitude of $V^\mathrm{nl}_\mathrm{\Delta}$, which however is not observed for the charge current densities $j_\mathrm{Cu}$ applied in this work.

\begin{figure}[!b]
	\centering
	\includegraphics[width=8.6 cm]{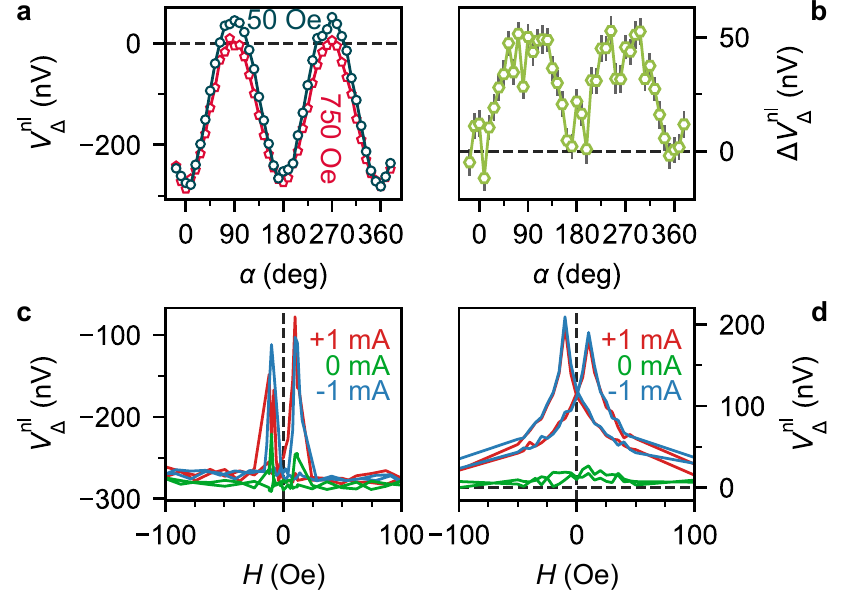}
	\caption{(b) Electrically induced non-local voltage $V^\mathrm{nl}_\mathrm{\Delta}$ [Pt~$\rightarrow$~Pt configuration] as a function of $\alpha$ and for different fields.
		(b) Signal difference $\Delta V^\mathrm{nl}_\mathrm{\Delta} = V^\mathrm{nl}_\mathrm{\Delta} \left(\SI{50}{Oe}\right) - V^\mathrm{nl}_\mathrm{\Delta} \left(\SI{750}{Oe}\right)$ as a function of $\alpha$, error bars are obtained by error propagation.
		(c),(d) Electrically ($V^\mathrm{nl}_\mathrm{\Delta}$) induced non-local voltage as a function of field (linear sweep) for different charge currents $J_\mathrm{Cu} = \SI{0}{\milli\ampere},\,\pm\SI{1}{\milli\ampere}$ applied to the Cu wire.
		In (c), positive fields align along $\alpha = \ang{0}$, while in (d) they show in the $\alpha = \ang{90}$ direction.
			Adapted from [\onlinecite{cramer2018phd}].}
	\label{fig:ptcupt7}
\end{figure}

Finally, the influence of the Oersted field generated by the Cu modulator on the spin transport signal was investigated.
As done before, angular-dependent measurements were repeated at high and low external magnetic fields, now with a fixed $J_\mathrm{Cu} = +\SI{1}{\milli\ampere}$.
In Fig.~\ref{fig:ptcupt7}a, $V^\mathrm{nl}_\mathrm{\Delta}$ is shown as a function of $\alpha$ and for two different external fields (red pentagons: \SI{750}{Oe}, blue circles: \SI{50}{Oe}).
The signals exhibit similar symmetries and for $\alpha = \ang{0}$/\ang{180}/\ang{360}, at which external and Oersted field are parallel, they furthermore have equal amplitudes.
At $\alpha = \ang{90}$ and $\alpha = \ang{270}$ (external and Oersted field perpendicular), however, $V^\mathrm{nl}_\mathrm{\Delta}$ deviates being zero at \SI{750}{Oe} and exhibiting a finite, positive value at \SI{50}{Oe}.
To highlight this discrepancy, the signal difference $\Delta V^\mathrm{nl}_\mathrm{\Delta} = V^\mathrm{nl}_\mathrm{\Delta} \left(\SI{50}{Oe}\right) - V^\mathrm{nl}_\mathrm{\Delta} \left(\SI{750}{Oe}\right)$ is shown in Fig.~\ref{fig:ptcupt7}b.
At first sight, this result might suggest that the spin (magnon) conductance of the YIG layer is altered by the Oersted field.
In a simple picture, $\bm{H}_\mathrm{Oe}$ cants $\bm{M}_\mathrm{YIG}$ away from the $\alpha = \ang{90}/\ang{270}$ direction, at which magnons cannot be injected or detected by the Pt wires due to the (I)SHE symmetry\cite{Sinova2015}.
If $\bm{H}_\mathrm{Oe}$ is strong enough to induce a finite $x$-component of $\bm{M}_\mathrm{YIG}$ underneath both injector and detector, these processes would become active again and a signal would appear.
However, the positive sign of $\Delta V^\mathrm{nl}_\mathrm{\Delta}$ in Fig.~\ref{fig:ptcupt7}b contradicts this interpretation.
For the realized device structure with equal injector and detector wire material, the electrically induced non-local voltage must be negative\cite{Goennenwein2015}.
SHE induced magnon flow may still be present, however the sign of $\Delta V^\mathrm{nl}_\mathrm{\Delta}$ implies a different, dominating effect that we discuss below.

To investigate the source of $\Delta V^\mathrm{nl}_\mathrm{\Delta}$, field and current sweep measurements were performed.
Regarding the field dependence, Fig.~\ref{fig:ptcupt7}c,d shows $V^\mathrm{nl}_\mathrm{\Delta}$ as a function of $H$ and $J_\mathrm{Cu}$ for $\alpha = \ang{0}$ (Fig.~\ref{fig:ptcupt7}c) and $\alpha = \ang{90}$ (Fig.~\ref{fig:ptcupt7}d).
The vertical solid lines mark the switching fields $H^\pm_\mathrm{c}$ of the YIG layer (see Supporting Information).
At $\alpha = \ang{0}$, the non-local voltage is mainly field-independent except for signal peaks near $H^\pm_\mathrm{c}$ (reduced absolute voltage).
These peaks are due to the YIG magnetization reversal and the associated formation of magnetic domains, whose random magnetization alignments impede the propagation of magnon spin currents.
For non-zero $J_\mathrm{Cu}$, the peaks are broader and of larger amplitude as compared to $J_\mathrm{Cu}= 0$, implying a more pronounced domain formation due to thermal activation or the presence of the Oersted field.
More specific information is revealed when considering $\alpha = \ang{90}$ in Fig~\ref{fig:ptcupt7}d.
In the case of $J_\mathrm{Cu} = \SI{0}{\milli\ampere}$,  $V^\mathrm{nl}_\mathrm{\Delta}$ remains zero except for a small positive deflection at low fields.
For $J_\mathrm{Cu} = \pm\SI{1}{\milli\ampere}$, however, a significant voltage signal appears, whose amplitude becomes largest at the coercive fields of the YIG layer.
Irrespective of the polarity of $J_\mathrm{Cu}$ and, hence, the direction of $\bm{H}_\mathrm{Oe}$, $V^\mathrm{nl}_\mathrm{\Delta}$ exhibits a positive sign so that it cannot be directly assigned to SSE currents generated underneath the Cu modulator (compare to previous results, e.g. Fig.~\ref{fig:ptcupt4}b).

\begin{figure}[!b]
	\centering
	\includegraphics[width=8.6 cm]{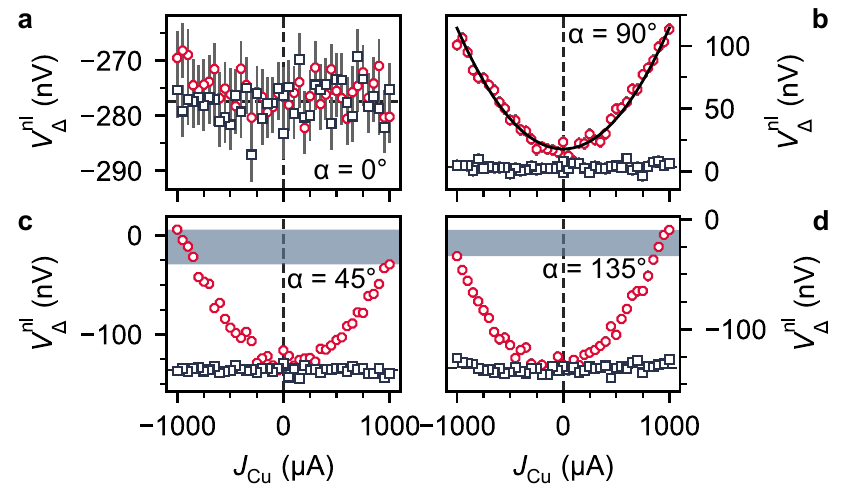}
	\caption{Current ($J_\mathrm{Cu}$) dependence of the electrically induced non-local voltage $V^\mathrm{nl}_\mathrm{\Delta}$ for different magnetization directions: (a) $\alpha = \ang{0}$, (b) $\alpha = \ang{90}$, (c) $\alpha = \ang{45}$ and (d) $\alpha = \ang{135}$.
		The results obtained for zero applied field (red circles) and an external field amplitude of $H = \SI{750}{Oe}$ (blue squares) are compared.
		The solid line in (b) corresponds to a quadratic fit function and the shaded areas in (c),(d) mark voltage level asymmetries for positive and negative $J_\mathrm{Cu}$.
		Error bars give the standard error.
			Adapted from [\onlinecite{cramer2018phd}].}
	\label{fig:ptcupt9}
\end{figure}

To understand this, the current ($J_\mathrm{Cu}$) dependence of $V^\mathrm{nl}_\mathrm{\Delta}$ is checked in Fig.~\ref{fig:ptcupt9} for different magnetization angles $\alpha$ and external fields (large vs. zero field).
Recall that $J_\mathrm{Cu}$ has no direct influence on the magnitude of the spin current excited electrically in the Pt injector.
Comparable to the results presented before (e.g. Fig.~\ref{fig:ptcupt2}b), $V^\mathrm{nl}_\mathrm{\Delta}$ exhibits no change with $J_\mathrm{Cu}$ at high fields (\SI{750}{Oe}, blue squares) for all magnetization directions.
This further holds true for zero field and $\alpha = \ang{0}$ (Fig.~\ref{fig:ptcupt9}a).
At $\alpha = \ang{90}$, however, $V^\mathrm{nl}_\mathrm{\Delta}$ exhibits a symmetric increase with current amplitude when no field is applied (red pentagons), see Fig.~\ref{fig:ptcupt9}b.
The data is fitted well by a quadratic function (solid line), which points towards a thermal origin.
For $\alpha = \ang{45}$ and $\alpha = \ang{135}$ (Fig.~\ref{fig:ptcupt9}c,d)  the zero-field signals grow with opposite asymmetry.

With this information, one can develop a model to explain the occurence of $\Delta V^\mathrm{nl}_\mathrm{\Delta}$ at low/zero fields and finite charge currents $J_\mathrm{Cu}$ applied to the Cu modulator.
At first, one has to consider that the Joule heat generated by $J_\mathrm{Cu}$ is not locally restricted but diffuses in the sample.
The phonon propagation length in YIG is of the order of several hundred micrometer \cite{Boona2014} so that the thermal equilibrium of the system is also strongly disturbed underneath the Pt injector (and detector).
Furthermore, the charge current $J_\mathrm{Pt}$ applied to the injector as well creates local Oersted fields.
One can thus conclude that the signal modulation $\Delta V^\mathrm{nl}_\mathrm{\Delta}$ is due to a difference between the $x$-components of thermally excited magnonic spin currents underneath the Pt injector.
It thus has the same origin as the non-local signal observed in the first part of this study (Cu $\rightarrow$ Pt, Fig.~\ref{fig:ptcupt2}b), which is corroborated by the similar symmtries of $V^\mathrm{nl}_\mathrm{\Delta}$ in Fig.~\ref{fig:ptcupt2}b and $\Delta V^\mathrm{nl}_\mathrm{\Delta}$ shown in Fig.~\ref{fig:ptcupt7}b.

In addition to the heat provided by the Cu modulator, the asymmetry of $V^\mathrm{nl}_\mathrm{\Delta}$ in Fig.~\ref{fig:ptcupt9}c,d signifies that the Oersted field generated by $J_\mathrm{Cu}$ as well influences the recorded signal.
Regarding first $\alpha = \ang{45}$, recall that $J_\mathrm{Cu} > 0$ induces a negative angular shift of the YIG magnetization, i.e., $\alpha^\prime = \ang{45} + \delta \alpha < \ang{45}$.
For $J_\mathrm{Cu} < 0$, we find $\alpha^\prime > \ang{45}$.
At the position of the Pt injector, this deflection of the YIG magnetization is further strengthened or weakened by the Oersted field induced by $J_\mathrm{Pt}$, depending on its polarity.
Now, based on the fact that $V^\mathrm{nl}_\mathrm{\Delta}$ is the voltage difference for positive and negative $J_\mathrm{Pt}$, not the amplitude but the slope of the original non-local SSE signal at $\alpha^\prime$ is decisive.
The latter follows a $\cos\left( \alpha^\prime\right)$ symmetry, such that
$$  \mathrm{abs}\left(\partial V^\mathrm{nl}_\mathrm{SSE} /\partial \alpha^\prime \right) \Bigr|_{\alpha^\prime = \ang{45}^-} < \left. \mathrm{abs}\left(\partial V^\mathrm{nl}_\mathrm{SSE} /\partial \alpha^\prime \right)\right| _{\alpha^\prime = \ang{45}^+}$$
holds.
One therefore expects a smaller $V^\mathrm{nl}_\mathrm{\Delta}$ for $J_\mathrm{Cu} > 0$ as compared to $J_\mathrm{Cu} < 0$, which agrees with the data shown in Fig.~\ref{fig:ptcupt9}c.
The inversed asymmetry for $\alpha = \ang{135}$ in Fig.~\ref{fig:ptcupt9}d can be explained by an analogous argumentation.

Altogether, these results demonstrate that the additional Joule heat and the Oersted fields provided by the Cu center wire indeed can be used to modulate the transport signal of electrically excited magnons.
As a final remark, note that YIG exhibits an exponentially decreasing electrical resistivity when exposed to strong resistive heating ($T \gg \SI{300}{\kelvin}$)\cite{Thiery2018a} such that electrically transmitted voltages may interfere with magnon mediated signals.
This effect may become important for large charge currents applied to the nanowires, nevertheless it cannot explain the distinct field and angular dependences observed in this work, which is why we can rule out charge transport effects as the dominating factor here.

\subsection{Conclusion}

In summary, the influence of localized heating and Oersted fields on magnonic spin transport signals in the insulating ferrimagnet YIG were investigated by using a non-local device structure with an additional Cu wire used to locally generate a field or induce a temperature change.
First experiments demonstrate that the exclusive application of a charge current to the Cu modulator generates a signal response, which exhibits similar features as SHE induced magnon flow.
The data reveals that this signal results from thermally excited magnons with different polarization.

A similar effect is observed when investigating the impact of the additional heat and Oersted fields on the spin transport signal between the outer Pt wires.
At small external fields, thermally induced magnon flow with different magnon polarization superimposes the conventional voltage response induced by electrically excited magnons.
A further modulation of the transport signal is observed at large external fields, for which the effect of $H_\mathrm{Oe}$ is suppressed.
This may be due to interference of the electrically excited magnon current with thermally activated magnons, however, further theoretical and experimental work (e.g. different non-local device geometries) is required to explain this observation quantitatively.

Overall, the results show that the magnon transport signal in a spin conduit such as YIG can be moduluated by localized heating and electromagnetic fields, which might find application in the field of magnon logic.

\section{Acknowledgments}
We kindly acknowledge support by the Deutsche Forschungsgemeinschaft (DFG) (SPP 1538 Spin Caloric Transport, SFB TRR173 SPIN+X in Mainz), the Graduate School of Excellence Materials Science in Mainz (DFG/GSC 266), and the EU project INSPIN (FP7-ICT-2013-X 612759).
L.B. acknowledges the EuropeanUnion’s Horizon 2020 research and innovation program underthe  Marie  Skłodowska-Curie  Grant  Agreement  ARTES  No.793159.  R.L.  acknowledges  the  European  Union’s  Horizon2020 research and innovation programme under the Marie Skłodowska-Curie Grant Agreement FAST No. 752195.

\bibliography{bibliography}

\end{document}